\documentclass[aps,prb,superscriptaddress,twocolumn,longbibliography]{revtex4-1}
\usepackage{bbm}
\usepackage{graphicx}% Include figure files
\usepackage{dcolumn}% Align table columns on decimal point
\usepackage{bm}% bold math
\usepackage{subfigure}
\usepackage{amsmath}

\usepackage{hyperref}
\hypersetup{
     colorlinks = true,
     citecolor  = magenta,  % cite
     linkcolor  = magenta % ref
}
\usepackage{attachfile}

\newcommand{\bk}{\boldsymbol k}

\newcommand{\zb}{\color {black}}

\usepackage{times}

\begin{document}

\title{Boundary flat bands with topological spin textures protected by sub-chiral symmetry}

\author{Yijie Mo}
\affiliation{Guangdong Provincial Key Laboratory of Magnetoelectric Physics and Devices, State Key Laboratory of Optoelectronic Materials and Technologies,
School of Physics, Sun Yat-sen University, Guangzhou 510275, China}

\author{Xiao-Jiao Wang}
\affiliation{Guangdong Provincial Key Laboratory of Magnetoelectric Physics and Devices, State Key Laboratory of Optoelectronic Materials and Technologies,
School of Physics, Sun Yat-sen University, Guangzhou 510275, China}

\author{Rui Yu}
\email{yurui@whu.edu.cn}
\affiliation{School of Physics and Technology, Wuhan University, Wuhan 430072, China}
\affiliation{Wuhan Institute of Quantum Technology, Wuhan 430206, China}

\author{Zhongbo Yan}
\email{yanzhb5@mail.sysu.edu.cn}
\affiliation{Guangdong Provincial Key Laboratory of Magnetoelectric Physics and Devices, State Key Laboratory of Optoelectronic Materials and Technologies,
School of Physics, Sun Yat-sen University, Guangzhou 510275, China}

\date{\today}

\begin{abstract}
Chiral symmetry plays an indispensable role in topological classifications as well as in the understanding of the origin of bulk or boundary flat bands.
The conventional definition of chiral symmetry refers to the existence of a constant unitary matrix anticommuting with the Hamiltonian.
As a constant unitary matrix has constant eigenvectors, boundary flat bands enforced by chiral symmetry, which share the same eigenvectors with the chiral symmetry operator, are dictated to carry fixed (pseudo)spin polarizations and be featureless in quantum geometry.
In this work, we generalize the chiral symmetry and introduce a concept termed sub-chiral symmetry.
Unlike the conventional chiral symmetry operator defined as constant {\zb matrix}, the sub-chiral symmetry operator depends on partial components of the momentum vector, so as its eigenvectors.
We show that topological gapped or gapless systems without chiral symmetry but with sub-chiral symmetry can support boundary flat bands, which exhibit topological spin textures and quantized Berry phases.
We expect that such intriguing boundary flat bands could give rise to a variety of exotic physics in the presence of interactions or disorders.
\end{abstract}

\maketitle

\section{Introduction}

The band theory is fundamental and powerful in the description of both quantum and classical periodic systems.
Being an exotic type of band structure, flat bands have triggered enduring and tremendous research interest in a diversity of  disciplines~\cite{Rafi2011,Bergholtz2013Review,Parameswaran2013,Liu2014review,Derzhko2015,Leykam2018Review,Rhim2021flat,Cao2018a,Cao2018b,Tang2020,Torma2022,Jiang2022review,Neupert2022review,Yin2022review}.
Due to the quench of kinetic energy, even weak interactions or disorders may have profound effects on a flat-band system, and this raises the possibility of the emergence of exotic correlated phases or peculiar transport phenomena. Well-known examples  include interaction-driven ferromagnetism~\cite{Mielke1991a,Mielke1993,Tasaki1998}, high-temperature superconductivity~\cite{Imada2000flat,Kopnin2011}, and disorder-driven inverse Anderson transition~\cite{Goda2006,Wang2022IAT,Zhang2023IAT}, to name a few.
When the flat bands carry nontrivial topology, it has been predicted that even more exotic correlated phases like fractional topological insulators with long-range entanglement can arise~\cite{Sun2011flat,Tang2011flat,Neupert2011flat}.

Flat bands generally imply the confined motion of electrons in real space, which may be induced by a confining potential or destructive interference effects associated with special lattice structures~\cite{Green2010flat,Week2012flat,Luis2016flat,Maimaiti2017flat,Maimaiti2019flat,Maimaiti2021flat,Rontgen2018,Pal2018flat,Mizoguchi2019flat,Chiu2020,Ma2020flat,Ogata2021flat,Liu2021flat,Morfonios2021flat,Mizoguchi2021flat,Hwang2021flat,Nakai2022}.
A textbook example of ideal flat bands is the Landau levels induced by a perpendicular magnetic field in two dimensions (2D)~\cite{Landau2013quantum}.
In this case, the magnetic field provides a potential to confine the motion of electrons. For translation-invariant systems, to have perfectly flat bands, in general, requires the existence of special symmetries to constrain the Hamiltonian.
The chiral symmetry is one such symmetry. Notably, the chiral symmetry can enable the realization of both bulk and boundary flat bands.
For instance, a bipartite lattice system with chiral symmetry and sublattice imbalance will have ideal bulk flat bands~\cite{Ramachandran2017flat}, with the number of flat bands precisely equal to the sublattice imbalance~\cite{Dumitru2022}.
Quite differently, the connection between chiral symmetry and boundary flat bands is through the bulk-boundary correspondence, a central property of topological phases~\cite{Chiu2015RMP}.
In chiral symmetric systems, a topological invariant known as winding number can be defined along 1D noncontractible loops in the Brillouin zone~\cite{Ryu2010}.
In dimensions $d\geq2$, a nonzero momentum-dependent winding number $\mathcal{W}(\bk_{\parallel})$ dictates the existence of $\mathcal{W}$ branches of zero-energy flat bands on each boundary with normal vector perpendicular to $\bk_{\parallel}$.
The 1D flat bands on the boundary of 2D Dirac semimetals/superconductors~\cite{Tanaka2010Andreev,Sato2011,Wang2013,Deng2014flat,Potter2014,Daido2017flat,Hu2018flat}, like the electronic flat bands on the zigzag edges in graphene~\cite{Nakada1996,neto2009} and the Andreev flat bands in $d$-wave high-temperature superconductors~\cite{Ryu2002}, and the 2D flat bands in 3D nodal-line semimetals/superconductors~\cite{Schnyder2011flat,Lu2015flat,Kopnin2011b,Burkov2011nodal,Yu2015,Kim2015,Fang2015,Chen2017link,Yan2017link,Bi2017knot} are celebrated examples of this class.

Despite being a generic guiding principle to realize boundary flat bands, the chiral symmetry to the boundary bands is like a double-edged sword.
On the one hand, it ensures the flatness and stability of the boundary bands.
On the other hand, it rules out the possibility of the presence of nontrivial quantum geometry and topology in the boundary bands.
The latter is because the zero-energy boundary states are also the eigenstates of
the chiral symmetry operator~\cite{Sato2011}, which itself is a constant unitary matrix.
{\zb As the chiral symmetry is an internal symmetry},
this fact implies that {\zb the cell-periodic part of the wave functions} of the boundary states are momentum-independent, so their derivatives, which determine the quantum geometry~\cite{Provost1980,Xiao2010review}, always vanish.
The constant {\zb cell-periodic part} also implies that the zero-energy states on a given boundary carry fixed (pseudo)spin polarizations.
In this work, we generalize the chiral symmetry and introduce a concept termed sub-chiral symmetry.
A fundamental difference between the two is that the symmetry operator of the sub-chiral symmetry is momentum-dependent and itself allows a topological characterization by the winding number like the Hamiltonian.
Due to this difference, the sub-chiral symmetry not only ensures the flatness and stability of the boundary-state bands but also allows the presence of nontrivial quantum geometry and topology in them.
Intriguingly, when the winding number characterizing the sub-chiral symmetry operator is nonzero, we find that the boundary flat bands carry both topological
spin textures and quantized Berry phases.

The structure of the paper is as follows. In Sec.\ref{sec2}, we establish the generic theory 
for the sub-chiral symmetry. In Sec.\ref{sec3}, we consider a two-band insulator model 
with sub-chiral symmetry and show that the
boundary flat bands carry topological spin textures and are characterized 
by a quantized $\pi$ Berry phase. In Sec.\ref{sec4}, we show that
similar physics can also appear in 3D and in topological semimetals. 
We provide more discussions on the experimental realization 
of systems with sub-chiral symmetry and conclude the paper in Sec.\ref{sec5}. 
Some calculation details are relegated to appendices.

\section{Generic theory\label{sec2}}

For a chiral symmetric Hamiltonian, there exists an operator $\mathcal{C}$ anticommuting with the Hamiltonian, i.e., $\{\mathcal{C},\mathcal{H}\}=0$.
The chiral symmetry is {\zb an internal} symmetry, so the chiral symmetry operator has no momentum dependence in any momentum-space basis~\cite{Chiu2015RMP}.
Besides the anticommutation relation with the Hamiltonian and the momentum independence, the chiral symmetry operator needs to satisfy two
more constraints, i.e., unitary and $\mathcal{C}^{2}=1$ (here $1$ denotes an identity matrix with dimension determined by the basis).
In this work, we generalize the chiral symmetry by only releasing the constraint of momentum independence.
As we will show below, such a generalization is justified and rather useful in understanding the properties of topological boundary states.

To be specific, when there exists a momentum-dependent unitary matrix satisfying
\begin{eqnarray}
\mathcal{C}(\bk_{\parallel})\mathcal{H}(k_{\perp},\bk_{\parallel})\mathcal{C}^{-1}(\bk_{\parallel})
=-\mathcal{H}(k_{\perp},\bk_{\parallel}),\label{gcs}
\end{eqnarray}
and $\mathcal{C}^{2}(\bk_{\parallel})=1$, we claim that the Hamiltonian $\mathcal{H}(k_{\perp},\bk_{\parallel})$ has a sub-chiral symmetry. The prefix ``sub'' describes the fact that the operator $\mathcal{C}(\bk_{\parallel})$ only depends on partial components of the momentum vector.
Here we have decomposed the momentum vector into two parts, i.e.,  $\bk=(k_{\perp},\bk_{\parallel})$.
The component $k_{\perp}$ refers to the momentum perpendicular to the edge or surface considered to be cut open, and $\bk_{\parallel}$ refers to the momentum components parallel to the edge or surface.
Why the sub-chiral symmetry can still be interpreted as a chiral symmetry is because when one focuses on the topological boundary states on a given boundary, the momenta along directions with periodic boundary conditions are good quantum numbers, and can be viewed as parameters of a 1D Hamiltonian.
When $\mathcal{C}(\bk_{\parallel})$ has no momentum dependence, it just goes back to the conventional chiral symmetry.

When the Hamiltonian has such a sub-chiral symmetry, a momentum-dependent winding number can accordingly be defined~\cite{Sato2011},
\begin{eqnarray}
\mathcal{W}(\bk_{\parallel})=\frac{1}{4\pi i}\oint dk_{\perp}\text{Tr}[\mathcal{C}(\bk_{\parallel})
\mathcal{H}^{-1}(\bk)\partial_{k_{\perp}}\mathcal{H}(\bk)].\label{wn}
\end{eqnarray}
This topological invariant counts the number of zero-energy states on a boundary and at the boundary momentum $\bk_{\parallel}$.

Notably, if the Hamiltonian satisfies Eq.(\ref{gcs}), there must exist a unitary operator $\mathcal{S}$ satisfying $\mathcal{S}^{2}=1$ and anticommuting with the sub-chiral symmetry operator (see Appendix \ref{Appendixb}), i.e.,
\begin{eqnarray}
\mathcal{S}\mathcal{C}(\bk_{\parallel})\mathcal{S}^{-1}=-\mathcal{C}(\bk_{\parallel}).
\end{eqnarray}
The above equation indicates that the sub-chiral symmetry operator itself has chiral symmetry, thereby one can further introduce a winding number to characterize the sub-chiral symmetry operator,
\begin{eqnarray}
\mathcal{W}_{c}=\frac{1}{4\pi i}\oint_{c} dk_{l}\text{Tr}[\mathcal{S}\mathcal{C}^{-1}(\bk_{\parallel})
\partial_{k_{l}}\mathcal{C}(\bk_{\parallel})],\label{dchiral}
\end{eqnarray}
where the integral is performed along a noncontractible or contractible loop in the boundary Brillouin zone.
A noncontractible loop refers to a momentum line traversing the boundary Brillouin zone.
As will be shown below, when $\mathcal{W}_{c}$ is a nonzero integer, the spin textures of boundary flat bands are also characterized by a winding number with its value equal to $\mathcal{W}_{c}$.
Furthermore, when $\mathcal{W}_{c}$ is an odd integer, we find that the boundary flat bands are characterized by a $\pi$ Berry phase.
Below we consider two explicit models in 2D and 3D to demonstrate the above generic physics.

\section{1D edge flat bands with topological spin textures\label{sec3}}

We first consider a tight-binding model in 2D,
\begin{eqnarray}
\mathcal{H}(\bk)&=&(m-t_{x}\cos k_{x}-t_{y}\cos k_{y})\sigma_{z}+\lambda_{2}\sin k_{x}\sin k_{y}\sigma_{y}\nonumber\\
&&+\lambda_{1}(\cos k_{y}+\delta)\sin k_{x}\sigma_{x},\label{2dHamiltonian}
\end{eqnarray}
where $\sigma_{x,y,z}$ are Pauli matrices, and $m$, $t_{x,y}$, $\lambda_{1,2}$ and $\delta$ are real parameters. Depending on the concrete physical realization, the Pauli matrices may act on either real spin or pseudo spin (e.g., orbitals).
To simplify the discussion, however, we do not emphasize their difference and always use spin to represent the two internal degrees of freedom.
In addition, all lattice constants are set to unity throughout for notational simplicity.

For the Hamiltonian in Eq.(\ref{2dHamiltonian}), it is easy to see that the chiral symmetry is absent as one cannot find a constant unitary operator to be anticommuting with the Hamiltonian.
According to the ten-fold way classification, this Hamiltonian belongs to the symmetry class AI as it only has the time-reversal symmetry, i.e., $\mathcal{T}\mathcal{H}(\bk) \mathcal{T}^{-1}=\mathcal{H}(-\bk)$, where$\mathcal{T}=\sigma_{z}\mathcal{K}$ with $\mathcal{K}$ being the complex conjugation operator, and $\mathcal{T}^{2}=1$.
In 2D, this symmetry class does not support any strong topological insulator phase~\cite{Schnyder2008,kitaev2009periodic}.
Nevertheless, this Hamiltonian turns out to have rather interesting bulk topology and boundary states.

Despite the absence of chiral symmetry, this Hamiltonian has the sub-chiral symmetry  defined in Eq.(\ref{gcs}), with the symmetry operator of the form
\begin{eqnarray}
\mathcal{C}(k_{y})=-\sin\theta(k_{y})\sigma_{x}+\cos\theta(k_{y})\sigma_{y},\label{gcsform}
\end{eqnarray}
where $\theta(k_{y})=\arg[\lambda_{1}(\cos k_{y}+\delta)+i\lambda_{2}\sin k_{y}]$.
Based on Eq.(\ref{wn}), a winding number can be defined,
\begin{eqnarray}
\mathcal{W}(k_{y})=\frac{1}{4\pi i}\int_{-\pi}^{\pi} dk_{x}\text{Tr}[\mathcal{C}(k_{y})
\mathcal{H}^{-1}(\bk)\partial_{k_{x}}\mathcal{H}(\bk)].
\end{eqnarray}
For the convenience of discussion, we assume $t_{x,y}$ and $\lambda_{1,2}$ to be positive.
We consider $t_{x}>|m-t_{y}\cos k_{y}|$ for arbitrary $k_{y}$ and
$\delta\neq\pm 1$ so that the Hamiltonian in Eq.(\ref{2dHamiltonian}) describes an insulator, then a straightforward calculation gives $\mathcal{W}(k_{y})=-1$ for arbitrary $k_{y}$.
For the sub-chiral symmetry operator given in Eq.(\ref{gcsform}), obviously, it anticommutes with $\sigma_{z}$, so its chiral symmetry operator is $\mathcal{S}=\sigma_{z}$. Using the formula in Eq.(\ref{dchiral}), one can find
\begin{eqnarray}
\mathcal{W}_{c}&=&\frac{1}{4\pi i}\int_{-\pi}^{\pi} dk_{y}\text{Tr}[\sigma_{z}\mathcal{C}^{-1}(k_{y})
\partial_{k_{y}}\mathcal{C}(k_{y})]\nonumber\\
&=&\frac{1}{2\pi}\int_{-\pi}^{\pi} dk_{y}\frac{\partial \theta(k_{y})}{\partial k_{y}}
=\left\{\begin{array}{cc}
            1, & |\delta|<1, \\
            0, & |\delta|>1.
          \end{array}\right.
\end{eqnarray}
The result indicates that the sub-chiral symmetry operator has a nontrivial winding in the regime $|\delta|<1$.

\begin{figure}[t]
\centering
\includegraphics[width=0.48\textwidth]{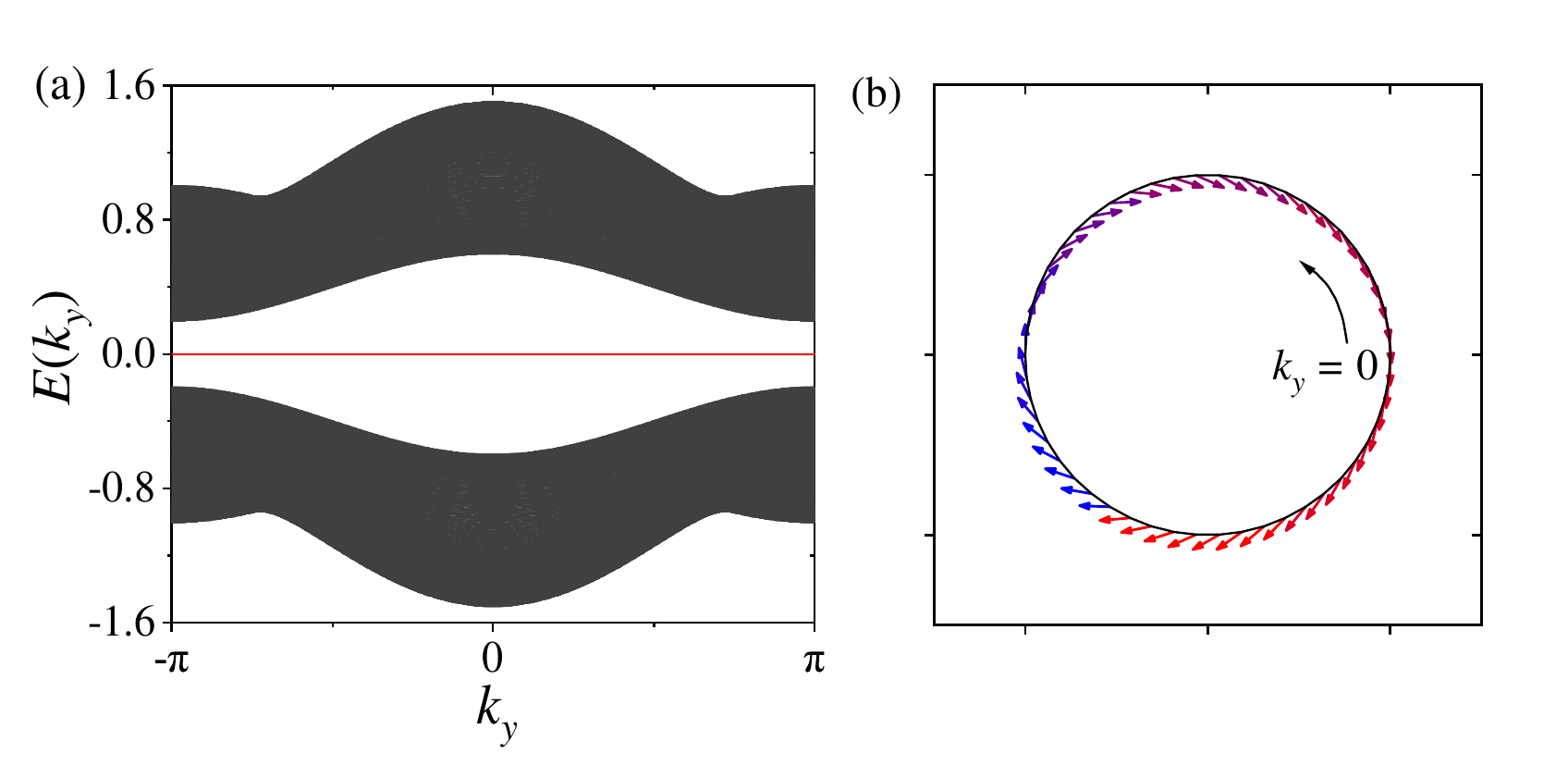}
\caption{(Color online) (a)
The solid red line of double degeneracy in the middle of the energy spectrum indicates the existence of one zero-energy flat band on each $x$-normal edge.
(b) Spin textures for the zero-energy flat band on the left $x$-normal edge.
Here the 1D boundary Brillouin zone is plotted as a cycle to show the winding of the spin textures better. Parameters are $m=0.2$, $t_x=0.6$, $t_y=0.2$, $\lambda_1=1$, $\lambda_2=1$, and $\delta=0.5$.
}\label{fig1}
\end{figure}

To show that a nonzero $\mathcal{W}(k_{y})$ ensures the existence of zero-energy flat bands, we choose a set of parameters leading to $\mathcal{W}(k_{y})=-1$ and consider a cylindrical sample with open (periodic) boundary conditions in the $x$ $(y)$ direction.
As shown in Fig.\ref{fig1}(a), the numerical result confirms the existence of one zero-energy flat band on each $x$-normal edge, verifying the correspondence with the winding number $\mathcal{W}(k_{y})$.

While a zero-energy flat band protected by chiral symmetry is dictated to carry a fixed spin polarization, below, we will show that the situation drastically changes for the zero-energy flat bands protected by sub-chiral symmetry.
To show this, we analytically extract the spin textures of the zero-energy edge flat bands exemplified in Fig.\ref{fig1}(a).
Here a key observation is that {\zb the cell-periodic part of the wave functions} of the zero-energy flat bands at a given boundary momentum must also be the eigenvector of the sub-chiral symmetry operator at the same momentum (see Appendix \ref{Appendixa}), which is reminiscent of the relation between zero-energy bound states and chiral symmetry operator.
The explicit steps are as follows.

First, according to Eq.(\ref{gcsform}), it is easy to find that the two eigenvectors of $\mathcal{C}(k_{y})$ are given by
\begin{eqnarray}
|\mathcal{C}(k_{y})=\pm1\rangle=\frac{1}{\sqrt{2}}
\left(\begin{array}{c}
                                                 1 \\
                                                 \pm ie^{i\theta(k_{y})}
                                               \end{array}\right).
\end{eqnarray}
Second, consider open boundary conditions in the $x$ direction.
Without loss of generality, we assume the lattice-site number in the $x$ direction to be $N$.
Accordingly, the Hamiltonian becomes a  $2N\times 2N$ matrix, and the form of the sub-chiral symmetry operator is expanded as
\begin{eqnarray}
\tilde{\mathcal{C}}(k_{y})=\mathbf{I}_{N}\otimes \mathcal{C}(k_{y}),
\end{eqnarray}
where $\mathbf{I}_{N}$ stands for the $N$-by-$N$ identity matrix.
The wave functions of the zero-energy states at the $x$-normal edges  will take the following general form (explicit expressions can be found in Appendix \ref{Appendixa})
\begin{eqnarray}
|\Psi_{\alpha}(k_{y})\rangle=(\xi_{1},\xi_{2},...,\xi_{N-1},\xi_{N})^{T}\otimes|\mathcal{C}(k_{y})=\beta\rangle,\label{WF}
\end{eqnarray}
where $\alpha$ labels left or right $x$-normal edge, and $\beta=-1$ (left edge) or $1$ (right edge).
$\xi_{i}$ {\zb characterizes the weight of the wave function at the $i$-th column of unit cells in the $x$ direction}, and the normalization condition requires $\sum_{i}|\xi_{i}|^{2}=1$.
Based on $|\Psi_{\alpha}\rangle$, one can determine the spin textures of the zero-energy edge flat bands by using the formula $\bar{\sigma}_{i}^{\alpha}(k_{y})=\langle \Psi_{\alpha}(k_{y})|\mathbf{I}_{N}\otimes \sigma_{i}|\Psi_{\alpha}(k_{y})\rangle$, which gives
\begin{eqnarray}
\bar{\sigma}_{x}^{\alpha}(k_{y})&=&-\beta\sin\theta(k_{y}), \nonumber\\
\bar{\sigma}_{y}^{\alpha}(k_{y})&=&\beta\cos\theta(k_{y}), \nonumber\\
\bar{\sigma}_{z}^{\alpha}(k_{y})&=&0.\label{spintexture}
\end{eqnarray}
Apparently, the spin textures will wind $n$ times around the origin if the argument $\theta(k_{y})$ changes $2n\pi$ when $k_{y}$ varies from $-\pi$ to $\pi$. In addition, the factor $\beta$ indicates that the spin textures on the two edges are just {\zb the} opposite.
In Fig.\ref{fig1}(b), we show the spin textures for one zero-energy edge flat band.
A complete cycle of winding in the spin polarizations is evident.

As the result in Eq.(\ref{spintexture}) reveals that the spin polarizations always lie in the $xy$ plane, a winding number can be further introduced to characterize the spin textures. Its form is
\begin{eqnarray}
\mathcal{W}_{s}^{\alpha}&=&\frac{1}{2\pi}\int_{-\pi}^{\pi} dk_{y}(\bar{\sigma}_{x}^{\alpha}\partial_{k_{y}}\bar{\sigma}_{y}^{\alpha}-
\bar{\sigma}_{y}^{\alpha}\partial_{k_{y}}\bar{\sigma}_{x}^{\alpha})=\mathcal{W}_{c}.\label{equal}
\end{eqnarray}
The above equation suggests that the nontrivial winding of the spin textures originates from the nontrivial winding of the sub-chiral symmetry operator.

Next, let us analyze the quantum geometry property of the zero-energy edge flat bands.
For zero-energy boundary flat bands protected by chiral symmetry, their Berry connections can always be made identically vanishing as
{\zb the cell-periodic part of the wave functions} has no momentum dependence. Hence, their Berry phases always take the trivial value, i.e., $\phi=0\mod 2\pi$.
In the current case, again the situation is drastically different.
Based on {\zb the cell-periodic part of the wave function in Eq.(\ref{WF})}, namely, $|\mathcal{C}(k_{y})=\beta\rangle$,
the Berry connections for the zero-energy edge flat bands are given by
\begin{eqnarray}
A_{\alpha}(k_{y})&=&-i\langle \mathcal{C}(k_{y})=\beta|\partial_{k_{y}}|\mathcal{C}(k_{y})=\beta\rangle\nonumber\\
&=&\frac{1}{2}\frac{\partial\theta(k_{y})}{\partial k_{y}}.
\end{eqnarray}
Immediately, one finds that the Berry phases associated with the two edge flat bands are~\cite{zak1989,Xiao2010review}
\begin{eqnarray}
\phi_{L}=\phi_{R}=\int_{-\pi}^{\pi}A_{L/R}(k_{y})dk_{y}=\mathcal{W}_{c}\pi \,(\text{mod}\, 2\pi).
\end{eqnarray}
The result indicates that, in the regime $\mathcal{W}_{c}=1$, the flat edge bands are characterized by a quantized $\pi$ Berry phase.

\section{2D surface flat bands with topological spin textures\label{sec4}}

Let us generalize the study to 3D and consider a topological semimetal to illustrate the generality of the physics. We consider the following simple model,
\begin{eqnarray}
\mathcal{H}(\bk)&=&(m-t\sum_{i=x,y,z}\cos k_{i})\sigma_{z}+\lambda\sin k_{z}\sin k_{x}\sin k_{y}\sigma_{y}\nonumber\\
&&+\lambda\sin k_{z}(\cos k_{x}-\cos k_{y})\sigma_{x}.
\end{eqnarray}
This Hamiltonian is the tight-binding counterpart of the low-energy continuum Hamiltonian developed by Xu {\it et al.} to describe the low-energy physics of a magnetic Weyl semimetal candidate HgCr$_{2}$Se$_{4}$~\cite{Xu2011Weyl}.
For the convenience of discussion, we again consider the parameters $t$ and $\lambda$ to be positive.
Due to the existence of a mirror symmetry $\mathcal{M}_{z}=i\sigma_{z}$, this Hamiltonian can support not only Weyl points but also nodal rings. Let us consider $t<m<3t$, which leads to the presence of two Weyl points at $\bk_{w}=\pm(0,0,\arccos((m-2t)/t))$
and a nodal ring in the $k_{z}=0$ plane. The nodal ring corresponds to the momentum contour satisfying $t(\cos k_{x}+\cos k_{y})=m-t$.
While the Weyl points and the concomitant Fermi arcs are of central interest in previous studies of this Hamiltonian~\cite{Xu2011Weyl,Fang2012weyl}, here we will focus on the zero-energy surface flat bands associated with the nodal ring.

\begin{figure}[t]
\centering
\includegraphics[width=0.48\textwidth]{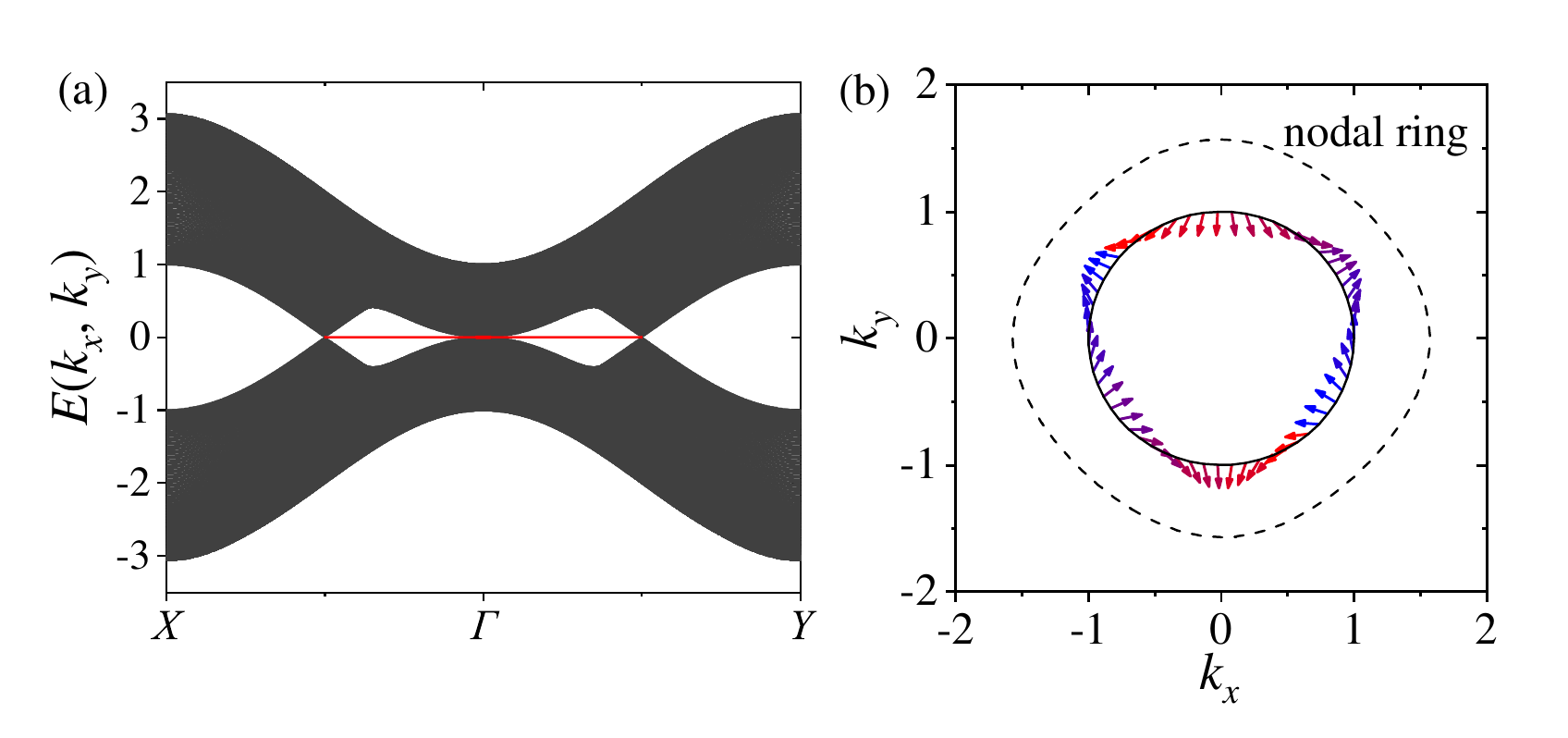}
\caption{(Color online) (a) Energy spectrum for a system with open (periodic) boundary
conditions in the $z$ ($x$ and $y$) direction.
The solid red line of double degeneracy indicates the existence of one zero-energy surface flat band
on each $z$-normal surface. $\Gamma=(0,0)$, $X=(\pi,0)$ and $Y=(0,\pi)$.
(b) Spin textures for the zero-energy flat band on the bottom $z$-normal surface. Within
the projection of the bulk nodal ring (dashed line), the in-plane spin polarizations on a circle wind two complete
cycles. Parameters are $m=2$, $t=1$, $\lambda=1$.
}\label{fig2}
\end{figure}

It is easy to see that this Hamiltonian also does not have
chiral symmetry, but has a sub-chiral symmetry with the symmetry operator
of the form
\begin{eqnarray}
\mathcal{C}(k_{x},k_{y})=-\sin \theta(k_{x},k_{y})\sigma_{x}+\cos \theta(k_{x},k_{y})\sigma_{y},\label{3dgcs}
\end{eqnarray}
where $\theta(k_{x},k_{y})=\arg[(\cos k_{x}-\cos k_{y})+i\sin k_{x}\sin k_{y}]$. Similarly,
following Eq.(\ref{wn}), one obtains
\begin{eqnarray}
\mathcal{W}(k_{x}, k_{y})=\left\{\begin{array}{cc}
                                   -1, & (k_{x},k_{y})\, \text{inside the nodal ring},\\
                                   0, & (k_{x},k_{y})\, \text{outside the nodal ring}.
                                 \end{array}
\right.
\end{eqnarray}
This indicates that when the $z$ direction is cut open,
zero-energy surface states exist only when the surface momentum
$(k_{x}, k_{y})$ falls inside the $z$-directional projection of the nodal ring.
In Fig.\ref{fig2}(a), the numerical result confirms that the zero-energy
surface bands do exist only in the mentioned region, verifying
the bulk-boundary correspondence.

Next, let us again first explore the topological property of the sub-chiral symmetry operator.
According to Eq.(\ref{3dgcs}), it is easy to find that its chiral symmetry operator $\mathcal{S}$ is also given by $\sigma_{z}$.
Applying Eq.(\ref{dchiral}), one obtains
\begin{eqnarray}
\mathcal{W}_{c}&=&\frac{1}{4\pi i}\oint_{c} dk_{l}\text{Tr}[\sigma_{z}\mathcal{C}^{-1}(k_{x},k_{y})
\partial_{k_{l}}\mathcal{C}(k_{x},k_{y})]=-2,\quad
\end{eqnarray}
where the integral contour is a loop enclosing the origin of the surface Brillouin zone
and falls inside the projection of the nodal ring. Here $\mathcal{W}_{c}=-2$ is simply because the argument $\theta(k_{x},k_{y})$ will change $4\pi$ when the polar angle of the momentum changes
$2\pi$.

Following the same analysis applied in 2D, one can find that the spin textures of the two zero-energy surface flat bands are given by
\begin{eqnarray}
\bar{\sigma}_{x}^{\alpha}(k_{x},k_{y})&=&-\beta\sin\theta(k_{x}, k_{y}), \nonumber\\
\bar{\sigma}_{y}^{\alpha}(k_{x},k_{y})&=&\beta\cos\theta(k_{x},k_{y}), \nonumber\\
\bar{\sigma}_{z}^{\alpha}(k_{x},k_{y})&=&0.
\end{eqnarray}
Here $\alpha$ labels the top and bottom $z$-normal surfaces,
and $\beta=1$ $(-1)$ for the top (bottom) surface, revealing that the spin textures on the top and bottom surfaces
are opposite.  Compared to $\theta(k_{y})$ in the two-dimensional model, here $\theta(k_{x}, k_{y})$ varies twice faster
along a loop around the origin,
corresponding to a faster variation of the spin textures, as shown in Fig.\ref{fig2}(b).
Using the formula for $\mathcal{W}_{s}^{\alpha}$ in Eq.(\ref{equal}),
it is easy to find $\mathcal{W}_{s}^{\alpha}=\mathcal{W}_{c}$, confirming again
the one-to-one correspondence between the winding of the spin textures and
the winding of the sub-chiral symmetry operator. For this specific model, as
$\mathcal{W}_{c}=-2$, the Berry phase
along a closed curve around the origin is thus trivial (mod $2\pi$).

\section{Discussions and conclusions\label{sec5}}

The generic theory and the two exemplary models above reveal that the sub-chiral symmetry is another generic symmetry principle to realize boundary flat bands.
As boundary flat bands protected by sub-chiral symmetry, unlike those protected by chiral symmetry, can carry topological spin textures and quantized Berry phases, they provide a more appealing basis to explore exotic phases driven by interaction or disorder effects.
About the experimental implementation of such boundary flat bands, we have
exemplified that they can appear in some topological quantum materials, {\zb like 3D nodal-line
semimetals}. {\zb Unconventional superconductors with appropriate pairings can also support
such topological boundary flat bands~\cite{Sigrist1991}} (see more discussions in Appendix \ref{Appendixc}).
Of course, they can also be easily implemented in artificial systems with even higher flexibility~\cite{Ma2019,Ozawa2019review,Lee2018,Wu2022circuit,Wang2023hopf}. {\zb In a follow-up
work, we have implemented the 2D Hamiltonian in a circuit system  and observed the predicted
boundary flat bands and concomitant topological spin textures as a proof of principle\cite{Biao2024}.}

In conclusion, we introduce the sub-chiral symmetry and reveal a class of boundary flat bands with fascinating properties.
Our study enriches the types of flat bands and hence opens a direction for the study of flat-band-related physics.

\section*{Acknowledgements}

Y. M., X.-J. W., and Z. Y. are supported by the National Natural Science Foundation of China
(Grant No. 12174455),  the Natural Science Foundation of Guangdong Province
(Grant No. 2021B1515020026), {\zb and the Guangdong Basic and Applied Basic Research Foundation (Grant No.
2023B1515040023)}.
R. Y. is supported by the National Natural Science Foundation of China (Grant No. 12274328), and the Beijing National Laboratory for Condensed Matter Physics.

\appendix

\section{Spin textures and quantized Berry phases of
the zero-energy edge flat bands in the two-dimensional model\label{Appendixa}}

The two-dimensional model is given by 
\begin{eqnarray}
\mathcal{H}(\bk)&=&(m-t_{x}\cos k_{x}-t_{y}\cos k_{y})\sigma_{z}+\lambda_{2}\sin k_{x}\sin k_{y}\sigma_{y}\nonumber\\
&&+\lambda_{1}(\cos k_{y}+\delta)\sin k_{x}\sigma_{x}.
\end{eqnarray}
To determine the spin textures and Berry phases of the zero-energy edge flat bands, we need to first determine the wave functions
of the zero-energy boundary states. Consider open boundary conditions in the $x$ direction and
periodic boundary conditions in the $y$ direction. Choosing the basis $\Psi=(c_{1,\uparrow,k_{y}},c_{1,\downarrow,k_{y}},...,
c_{N,\uparrow,k_{y}},c_{N,\downarrow,k_{y}})^{T}$, the Hamiltonian is accordingly given by
\begin{widetext}
\begin{eqnarray}
H=\left(
    \begin{array}{ccccccc}
      m(k_{y}) & 0 & -\frac{t_{x}}{2} & -i\frac{\Lambda(k_{y})e^{-i\theta(k_{y})}}{2} & 0 & 0 & \cdots \\
      0 & -m(k_{y}) & -i\frac{\Lambda(k_{y})e^{i\theta(k_{y})}}{2} & \frac{t_{x}}{2} & 0 & 0 & \cdots \\
      -\frac{t_{x}}{2} & i\frac{\Lambda(k_{y})e^{-i\theta(k_{y})}}{2} & m(k_{y}) & 0 & -\frac{t_{x}}{2} & -i\frac{\Lambda(k_{y})e^{-i\theta(k_{y})}}{2} & \cdots \\
      i\frac{\Lambda(k_{y})e^{i\theta(k_{y})}}{2} & \frac{t_{x}}{2} & 0 & -m(k_{y}) & -i\frac{\Lambda(k_{y})e^{i\theta(k_{y})}}{2} & \frac{t_{x}}{2} & \cdots \\
      0 & 0 & -\frac{t_{x}}{2} & i\frac{\Lambda(k_{y})e^{-i\theta(k_{y})}}{2} & m(k_{y}) & 0 & \cdots \\
      0 & 0 & i\frac{\Lambda(k_{y})e^{i\theta(k_{y})}}{2} & \frac{t_{x}}{2} & 0 & -m(k_{y}) & \cdots \\
      \vdots & \vdots & \vdots & \vdots & \vdots & \vdots & \ddots \\
    \end{array}
  \right),
\end{eqnarray}
\end{widetext}
where $m(k_{y})=m-t_{y}\cos k_{y}$, $\Lambda(k_{y})=\sqrt{\lambda_{1}^{2}(\cos k_{y}+\delta)^{2}+\lambda_{2}^{2}\sin^{2}k_{y}}$,
and $\theta(k_{y})=\arg[\lambda_{1}(\cos k_{y}+\delta)+i\lambda_{2}\sin k_{y}]$.

Solving the eigenvalue equation
\begin{eqnarray}
H|\Psi\rangle=E|\Psi\rangle,
\end{eqnarray}
where
\begin{eqnarray}
|\Psi\rangle=(\psi_{1\uparrow},\psi_{1\downarrow},...,\psi_{N\uparrow},\psi_{N\downarrow})^{T}.
\end{eqnarray}
For zero-energy eigenstates, one can find that they satisfy the following iterative equations,
\begin{widetext}
\begin{eqnarray}
&m(k_{y})\psi_{1\uparrow}-\frac{t_{x}}{2}\psi_{2\uparrow}-i\frac{\Lambda(k_{y})e^{-i\theta(k_{y})}}{2}\psi_{2\downarrow}=0,\nonumber\\
&-m(k_{y})\psi_{1\downarrow}-i\frac{\Lambda(k_{y})e^{i\theta(k_{y})}}{2}\psi_{2\uparrow}+\frac{t_{x}}{2}\psi_{2\downarrow}=0,\nonumber\\
&-\frac{t_{x}}{2}\psi_{1\uparrow}+i\frac{\Lambda(k_{y}) e^{-i\theta(k_{y})}}{2}\psi_{1\downarrow}+m(k_{y})\psi_{2\uparrow}
-\frac{t_{x}}{2}\psi_{3\uparrow}-i\frac{\Lambda(k_{y}) e^{-i\theta(k_{y})}}{2}\psi_{3\downarrow}=0,\nonumber\\
&i\frac{\Lambda(k_{y}) e^{i\theta(k_{y})}}{2}\psi_{1\uparrow}+\frac{t_{x}}{2}\psi_{1\downarrow}-m(k_{y})\psi_{2\downarrow}
-i\frac{\Lambda(k_{y})e^{i\theta(k_{y})}}{2}\psi_{3\uparrow}+\frac{t_{x}}{2}\psi_{3\downarrow}=0,\nonumber\\
&...\nonumber\\
&-\frac{t_{x}}{2}\psi_{n-1\uparrow}+i\frac{\Lambda(k_{y}) e^{-i\theta(k_{y})}}{2}\psi_{n-1\downarrow}+m(k_{y})\psi_{n\uparrow}
-\frac{t_{x}}{2}\psi_{n+1\uparrow}-i\frac{\Lambda(k_{y}) e^{-i\theta(k_{y})}}{2}\psi_{n+1\downarrow}=0,\nonumber\\
&i\frac{\Lambda(k_{y}) e^{i\theta(k_{y})}}{2}\psi_{n-1\uparrow}+\frac{t_{x}}{2}\psi_{n-1\downarrow}-m(k_{y})\psi_{n\downarrow}
-i\frac{\Lambda(k_{y})e^{i\theta(k_{y})}}{2}\psi_{n+1\uparrow}+\frac{t_{x}}{2}\psi_{n+1\downarrow}=0,....\label{iteration}
\end{eqnarray}
\end{widetext}
Since the Hamiltonian has the sub-chiral symmetry, the zero-energy states on the boundary
must be the eigenstates of the sub-chiral symmetry operator. This fact can simply be inferred by noting
that if $|\Psi\rangle$ is the eigen wave function of a zero-energy state, i.e., $H|\Psi\rangle=0$, then
as
\begin{eqnarray}
H\mathcal{C}(k_{y})|\Psi\rangle=-\mathcal{C}(k_{y})H|\Psi\rangle=0,
\end{eqnarray}
and $\mathcal{C}(k_{y})$ is an on-site operator for a given $k_{y}$ (the good quantum number $k_{y}$ can be treated
as a parameter when considering the boundary states on the $x$-normal edges),
$\mathcal{C}(k_{y})|\Psi\rangle$ must be equal to $|\Psi\rangle$ up to a phase factor. On the other hand,
since $\mathcal{C}(k_{y})^{2}=1$, one obtains
\begin{eqnarray}
\mathcal{C}(k_{y})|\Psi\rangle=|\Psi\rangle, \quad\text{or}\quad \mathcal{C}(k_{y})|\Psi\rangle=-|\Psi\rangle,
\end{eqnarray}
confirming that the eigen wave function of a zero-energy state is also the eigen wave function of the sub-chiral
symmetry operator.

As the sub-chiral symmetry operator takes
the form $\mathcal{C}(k_{y})=-\sin\theta(k_{y})\sigma_{x}+\cos\theta(k_{y})\sigma_{y}$, its two eigenvectors
read
\begin{eqnarray}
|\mathcal{C}(k_{y})=1\rangle&=&\frac{1}{\sqrt{2}}
\left(\begin{array}{c}
                                                 1 \\
                                                 ie^{i\theta(k_{y})}
                                               \end{array}\right),\nonumber\\
|\mathcal{C}(k_{y})=-1\rangle&=&\frac{1}{\sqrt{2}}
\left(\begin{array}{c}
                                                 1 \\
                                                 -ie^{i\theta(k_{y})}
                                               \end{array}\right),
\end{eqnarray}
where $|\mathcal{C}(k_{y})=\pm1\rangle$ satisfy $\mathcal{C}(k_{y})|\mathcal{C}(k_{y})=\pm1\rangle=\pm|\mathcal{C}(k_{y})=\pm1\rangle$.
Accordingly, for the zero-energy states, the spinor $(\psi_{j\uparrow},\psi_{j\downarrow})^{T}$ for each unit cell
must be proportional to either $|\mathcal{C}(k_{y})=1\rangle$ or $|\mathcal{C}(k_{y})=-1\rangle$. Let us focus on the zero-energy
states on the left $x$-normal edge. If one considers the special limit $m(k_{y})=0$ and $\Lambda(k_{y})=t_{x}$, it is easy
to see that the wave function for the zero-energy state is of the simple form
\begin{eqnarray}
|\Psi_{L}\rangle=\frac{1}{\sqrt{2}}(1,-ie^{i\theta(k_{y})},0,0,0,0,...)^{T}.
\end{eqnarray}
Based on this special case, one knows that for the zero-energy state on the left $x$-normal edge,
the spinor $(\psi_{j\uparrow},\psi_{j\downarrow})^{T}$ is proportional to $|\mathcal{C}(k_{y})=-1\rangle$.
With this observation, for the more generic case, we can assume that the
wave function takes the following trial form
\begin{eqnarray}
\psi_{j}=\left(\begin{array}{c}
  \psi_{j\uparrow} \\
  \psi_{j\downarrow}
\end{array}\right)=C\xi^{j}\left(\begin{array}{c}
                                                 1 \\
                                                 -ie^{i\theta(k_{y})}
                                               \end{array}\right)/\sqrt{2},
\end{eqnarray}
where $|\xi|<1$ must be enforced to ensure the decaying nature of the wave function of the zero-energy bound
state, and $C$ is a constant for later normalization. Taking the above expression into the series of iterative
equations, one can find that they become
\begin{eqnarray}
(-\frac{t_{x}}{2}+\frac{\Lambda(k_{y})}{2})\xi^{j-1}+m(k_{y})\xi^{j}-(\frac{t_{x}}{2}+\frac{\Lambda(k_{y})}{2})\xi^{j+1}=0\nonumber\\
\end{eqnarray}
for $j\geq2$, which can be further reduced to the following equation,
\begin{eqnarray}
(t_{x}+\Lambda(k_{y}))\xi^{2}-2m(k_{y})\xi+(t_{x}-\Lambda(k_{y}))=0.
\end{eqnarray}
The solutions are
\begin{eqnarray}
\xi_{\pm}=\frac{m(k_{y})\pm\sqrt{m^{2}(k_{y})+\Lambda^{2}(k_{y})-t_{x}^{2}}}{t_{x}+\Lambda(k_{y})}.
\end{eqnarray}
As $\Lambda(k_{y})>0$, one can verify $|\xi_{\pm}|<1$ as long as $|m(k_{y})|<t_{x}$,
which is consistent with the bulk-boundary correspondence, which says
that when the bulk topological invariant $\mathcal{W}(k_{y})=-1$, each $x$-normal edge
will harbor one zero-energy bound state.

Because of the existence of two solutions for $\xi$, the wave function for the zero-energy state
will take the form
\begin{eqnarray}
\left(\begin{array}{c}
  \psi_{j\uparrow} \\
  \psi_{j\downarrow}
\end{array}\right)=\mathcal{N}(C_{+}\xi_{+}^{j}+C_{-}\xi_{-}^{j})\left(\begin{array}{c}
                                                 1 \\
                                                 -ie^{i\theta(k_{y})}
                                               \end{array}\right)/\sqrt{2},\nonumber\\
\end{eqnarray}
where $\mathcal{N}$ stands for the normalization constant.
Enforcing the physical boundary condition $\psi_{j\uparrow}=\psi_{j\downarrow}=0$ for $j=0$, one obtains
$C_{+}=-C_{-}=C$. Bringing the above expression into the first two equations in the series shown
in Eq.(\ref{iteration}), one gets
\begin{eqnarray}
C(\xi_{+}-\xi_{-})m(k_{y})-\frac{(t_{x}+\Lambda(k_{y}))}{2}C(\xi_{+}^{2}-\xi_{-}^{2})=0.\nonumber\\
\end{eqnarray}
It is easy to find that the equation is naturally satisfied regardless of the value of $C$. Therefore,
we can set $C=1$. Accordingly, we obtain
\begin{eqnarray}
\left(\begin{array}{c}
  \psi_{j\uparrow} \\
  \psi_{j\downarrow}
\end{array}\right)=\mathcal{N}(\xi_{+}^{j}-\xi_{-}^{j})\left(\begin{array}{c}
                                                 1 \\
                                                 -ie^{i\theta(k_{y})}
                                               \end{array}\right)/\sqrt{2}.
\end{eqnarray}
Before proceeding, it is worth pointing out that $\xi_{+}$ and $\xi_{-}$ are either both
real or complex conjugate to each other. This fact implies that the spatial part of the
wave function can always be made real, a property that will be used to derive the
Berry connection and Berry phase.

Using the normalization condition,
\begin{eqnarray}
&&\sum_{j=0}^{\infty}(|\psi_{j\uparrow}|^{2}+|\psi_{j\downarrow}|^{2})\nonumber\\
&=&\mathcal{N}^{2}[|\xi_{+}-\xi_{-}|^{2}+|\xi_{+}^{2}-\xi_{-}^{2}|^{2}+...+|\xi_{+}^{n}-\xi_{-}^{n}|^{2}+...]\nonumber\\
&=&\mathcal{N}^{2}[\sum_{n=1}|\xi|_{+}^{2n}-\sum_{n=1}(\xi_{+}^{*}\xi_{-})^{n}-\sum_{n=1}(\xi_{+}\xi_{-}^{*})^{n}+\sum_{n=1}|\xi_{-}|^{2n}]\nonumber\\
&=&\mathcal{N}^{2}[\frac{|\xi_{+}|^{2}}{1-|\xi_{+}|^{2}}+\frac{|\xi_{-}|^{2}}{1-|\xi_{-}|^{2}}-\frac{\xi_{+}^{*}\xi_{-}}{1-\xi_{+}^{*}\xi_{-}}
-\frac{\xi_{+}\xi_{-}^{*}}{1-\xi_{+}\xi_{-}^{*}}]\nonumber\\
&=&1,
\end{eqnarray}
one determines
\begin{eqnarray}
\mathcal{N}=\frac{1}{\sqrt{[\frac{|\xi_{+}|^{2}}{1-|\xi_{+}|^{2}}+\frac{|\xi_{-}|^{2}}{1-|\xi_{-}|^{2}}-\frac{\xi_{+}^{*}\xi_{-}}{1-\xi_{+}^{*}\xi_{-}}
-\frac{\xi_{+}\xi_{-}^{*}}{1-\xi_{+}\xi_{-}^{*}}]}}.\quad
\end{eqnarray}

{\zb Let us now move to calculate the Berry connection and Berry phase.
As is known, for Bloch states, the Berry connection is determined by the cell-periodic part of the Bloch wave functions. For
here the boundary states, their wave functions can also be interpreted as Bloch states, but
the associated momenta in the plane-wave part are complex rather than real due to the decaying
nature of the wave functions of the boundary states. Based on this perspective, the
Berry connection for the boundary flat bands should also be determined by the cell-periodic part.
Accordingly, we have
\begin{eqnarray}
A_{L}(k_{y})&=&-i\langle \mathcal{C}(k_{y})=-1|\partial_{k_{y}}|\mathcal{C}(k_{y})=-1\rangle\nonumber\\
&=&\frac{1}{2}\frac{\partial \theta(k_{y})}{\partial k_{y}}.
\end{eqnarray}
This is one perspective. We can also adopt another different perspective to define the Berry connection.
That is, we can view that all lattice sites in the open boundary direction (here the $x$ direction, which is
the decaying direction of the boundary states) as
sublattice degrees of freedom of a big unit cell. Under this perspective, the two-dimensional Hamiltonian
effectively reduces to a one-dimensional Hamiltonian. Since now all lattice sites in the $x$ direction
become internal degrees of freedom, the cell-periodic part of these zero-energy states become $|\Psi_{L}(k_{y})\rangle$.
Accordingly,  the Berry connection is defined as
\begin{eqnarray}
A_{L}(k_{y})&=&-i\langle \Psi_{L}(k_{y})|\partial_{k_{y}}\Psi_{L}(k_{y})\rangle.
\end{eqnarray}
The above two definitions of the Berry connection are in fact equivalent.
To see this, note that a direct calculation of the above formula gives
\begin{widetext}
\begin{eqnarray}
A_{L}(k_{y})&=&-i\langle \Psi_{L}(k_{y})|\partial_{k_{y}}\Psi_{L}(k_{y})\rangle\nonumber\\
&=&\frac{1}{2}\frac{\partial \theta(k_{y})}{\partial k_{y}}-i(\mathcal{N}^{2}\sum_{n=1}(\xi_{+}^{n*}-\xi_{-}^{n*})n(\xi_{+}^{n-1}
\frac{\partial \xi_{+}}{\partial k_{y}}-\xi_{-}^{n-1}\frac{\partial \xi_{-}}{\partial k_{y}})+\mathcal{N}\frac{\partial \mathcal{N}}{\partial k_{y}}\sum_{n=1}|\xi_{+}^{n}-\xi_{-}^{n}|^{2}).
\end{eqnarray}
\end{widetext}
The next step is to show that, in the second line of the above equation,  the terms in the bracket must cancel with each other.
This fact can easily be inferred by recalling that $\xi_{+}$ and $\xi_{-}$ are either both
real or complex conjugate to each other. When $\xi_{+}$ and $\xi_{-}$ are real (when $\Lambda(k_{y})>t_{x}$ for arbitrary $k_{y}$, the realness
of $\xi_{+}$ and $\xi_{-}$ is ensured),
the terms in the bracket must cancel with each other as
the intraband Berry connection is a real quantity. The other situation
is that $\xi_{+}$ and $\xi_{-}$ are complex and conjugated to each other,
i.e., $\xi_{+}^{*}=\xi_{-}$. It is easy to see that, for this situation,
the two terms in the bracket are also real, so they also have to cancel with
each other due to the same reason. }

Now we are going to show that the zero-energy flat band on the left $x$-normal edge is characterized
by a quantized Berry phase. In terms of the defined Berry connection, it is straightforward to find
\begin{eqnarray}
\phi_{L}&=&\int_{-\pi}^{\pi}A_{L}(k_{y})dk_{y}=\frac{1}{2}[\theta(\pi)-\theta(-\pi)]\nonumber\\
&=&\left\{
                                                                                                      \begin{array}{cc}
                                                                                                        \pi \mod 2\pi, & |\delta|<1, \\
                                                                                                        0 \mod 2\pi, & |\delta|>1. \\
                                                                                                      \end{array}
                                                                                                   \right.
\end{eqnarray}
Similar analysis reveals that the zero-energy flat band on the right $x$-normal edge
is also characterized by a quantized Berry phase, and
\begin{eqnarray}
\phi_{R}&=&\int_{-\pi}^{\pi}A_{R}(k_{y})dk_{y}\nonumber\\
&=&-i\int_{-\pi}^{\pi}\langle \mathcal{C}(k_{y})=1|\partial_{k_{y}}\mathcal{C}(k_{y})=1\rangle\nonumber dk_{y}\\
&=&\frac{1}{2}\int_{-\pi}^{\pi}\frac{\partial\theta(k_{y})}{\partial k_{y}} dk_{y}\nonumber\\
&=&\frac{1}{2}[\theta(\pi)-\theta(-\pi)]=\phi_{L}.
\end{eqnarray}

For the spin textures of the zero-energy flat band on the left $x$-normal edge,
it is easy to find out that
\begin{eqnarray}
\bar{\sigma}_{x,L}&=&\sum_{j=1}\psi_{j}^{\dag}\sigma_{x}\psi_{j}=\sin \theta(k_{y}),\nonumber\\
\bar{\sigma}_{y,L}&=&\sum_{j=1}\psi_{j}^{\dag}\sigma_{y}\psi_{j}=-\cos \theta(k_{y}),\nonumber\\
\bar{\sigma}_{z,L}&=&\sum_{j=1}\psi_{j}^{\dag}\sigma_{z}\psi_{j}=0.
\end{eqnarray}
Similar analysis shows that the spin textures of the zero-energy flat band on the right $x$-normal edge
are just opposite, i.e.,
\begin{eqnarray}
\bar{\sigma}_{x,R}&=&-\sin \theta(k_{y}),\nonumber\\
\bar{\sigma}_{y,R}&=&\cos \theta(k_{y}),\nonumber\\
\bar{\sigma}_{z,R}&=&0.
\end{eqnarray}

The derivation for the three-dimensional model is similar, and hence we don't repeat
the details.

\section{Chiral symmetry of the sub-chiral symmetry operator\label{Appendixb}}

In the main text, we have stated that when the Hamiltonian has sub-chiral symmetry, the sub-chiral symmetry operator
itself has chiral symmetry. In this section, we demonstrate this fact explicitly.

As we consider Hamiltonians without chiral symmetry but with sub-chiral symmetry, the general forms of the Hamiltonians can
be expressed as
\begin{eqnarray}
\mathcal{H}(\bk)=\sum_{i=1}^{2n+1}d_{i}(\bk)\gamma_{i},
\end{eqnarray}
where the Hamiltonian is a $2n\times 2n$ matrix with $n$ a nonzero positive integer, $d_{i}(\bk)$ are real functions of the momentum,
and $\gamma_{i}$ are matrices satisfying the Clifford algebra, i.e., $\{\gamma_{i},\gamma_{j}\}=2\delta_{ij}\mathbf{I}_{2n}$
with $\mathbf{I}_{2n}$ the $2n\times 2n$ identity matrix. It is noteworthy that, as the number of $2n\times 2n$ matrices
satisfying the Clifford algebra is just $2n+1$, one cannot find an additional constant matrix to identically anticommute with the Hamiltonian if none of the
$d_{i}(\bk)$ is identically zero and no more than one is a constant (if two or more of them are a momentum-independent constant, one can just
rotate the basis and remove at least one of the $\gamma$ matrices, and the removed one will anticommute with the Hamiltonian, which is in contradiction with the absence of chiral symmetry).
Here we consider that all  $d_{i}(\bk)$ are dispersive functions of the momentum, so that the chiral
symmetry is dictated to be absent.

Let us now discuss the sub-chiral symmetry operator. As this operator anticommutes with the Hamiltonian
and satisfies $\mathcal{C}^{2}(\bk_{\parallel})=\mathbf{I}_{2n}$ (here $\bk_{\parallel}$ denotes
the set of momentum components that can be treated as good quantum numbers for the concerned problem),
its form can also be expressed in terms of
the $\gamma$ matrices, i.e.,
\begin{eqnarray}
\mathcal{C}(\bk_{\parallel})=\sum_{i=1}^{2n+1}c_{i}(\bk_{\parallel})\gamma_{i},\,\text{with}\,\sum_{i=1}^{2n+1}c_{i}^{2}(\bk_{\parallel})=1.
\label{normalization}
\end{eqnarray}
Now we are going to show that if the Hamiltonian has sub-chiral symmetry, then at least one of the $\gamma$ matrices  can be removed
from the above generic form of the sub-chiral symmetry operator. To see this, it is instructive to show that the existence of sub-chiral symmetry,
which requires $\{\mathcal{C}(\bk_{\parallel}), \mathcal{H}(\bk)\}=0$, leads to the following equation,
\begin{eqnarray}
\sum_{i=1}^{2n+1}c_{i}(\bk_{\parallel})d_{i}(\bk)=0.\label{chiral}
\end{eqnarray}
With the additional normalization condition shown in Eq.(\ref{normalization}), the existence of sub-chiral symmetry thus
only needs to satisfy two conditions under the constraint that the sub-chiral symmetry operator
only depends on partial components of the momentum. Apparently, the condition in Eq.(\ref{chiral}) can always be fulfilled if the ratio
of some (at least two) of the $d_{i}(\bk)$ only depends on $\bk_{\parallel}$.
Without loss of generality, let us assume that the ratio between $d_{1}(\bk)$ and $d_{2}(\bk)$, $\eta(\bk)=d_{1}(\bk)/d_{2}(\bk)$,
is a function of $\bk_{\parallel}$ only, i.e., $\eta(\bk)=\eta(\bk_{\parallel})$. Then apparently the sub-chiral symmetry operator
can be chosen as the form,
\begin{eqnarray}
\mathcal{C}(\bk_{\parallel})=\frac{1}{\sqrt{1+\eta^{2}(\bk_{\parallel})}}[-\gamma_{1}+\eta(\bk_{\parallel})\gamma_{2}].
\end{eqnarray}
Since the number of $\gamma$ matrices is $2n+1$ with $n\geq1$, there always exist at least
one $\gamma$ matrix to anticommute with the sub-chiral symmetry operator, verifying
that the sub-chiral symmetry operator itself has chiral symmetry.  It is worth mentioning
that, for a given set of partial components of the momentum, if there are only two $d$-functions
whose ratio depends on these momentum components, then the form of the sub-chiral symmetry operator is
definite up to an unimportant sign (multiply a minus sign to $\mathcal{C}(\bk_{\parallel})$ does not affect
the anticommutation relation $\{\mathcal{C}(\bk_{\parallel}), \mathcal{H}(\bk)\}=0$).
The two models studied in the paper belong to such a case.

When there exists more than two $d(\bk)$-functions whose mutual ratios depend on
the same set of momentum components, then the form of the sub-chiral symmetry operator
is not unique and in fact has infinitely many choices. To illustrate this fact intuitively, let
us consider a concrete example. Without loss of
generality, let us consider the simplest case for which the mutual ratios among $d_{1}(\bk)$, $d_{2}(\bk)$ and $d_{3}(\bk)$
all only depend on the same set of momentum components, say $\bk_{\parallel}$. To be specific, let us define
$\eta_{1}(\bk_{\parallel})=d_{1}(\bk)/d_{2}(\bk)$ and $\eta_{2}(\bk_{\parallel})=d_{1}(\bk)/d_{3}(\bk)$. It is
easy to check that the following two choices,
\begin{eqnarray}
\mathcal{C}_{1}(\bk_{\parallel})&=&\frac{1}{\sqrt{1+\eta_{1}^{2}(\bk_{\parallel})}}[-\gamma_{1}+\eta_{1}(\bk_{\parallel})\gamma_{2}],\nonumber\\
\mathcal{C}_{2}(\bk_{\parallel})&=&\frac{1}{\sqrt{1+\eta_{2}^{2}(\bk_{\parallel})}}[-\gamma_{1}+\eta_{2}(\bk_{\parallel})\gamma_{3}],
\end{eqnarray}
both satisfy the anticommutation relation with the Hamiltonian, i.e.,  $\{\mathcal{C}_{1}(\bk_{\parallel}), \mathcal{H}(\bk)\}=\{\mathcal{C}_{2}(\bk_{\parallel}), \mathcal{H}(\bk)\}=0$. As $\{\gamma_{i},\gamma_{j}\}=2\delta_{ij}\mathbf{I}_{2n}$,
one can view $\mathcal{C}_{1}(\bk_{\parallel})$ and $\mathcal{C}_{2}(\bk_{\parallel})$ as two unit vectors in the (2n+1)-dimensional space
spanned by the (2n+1) $\gamma$ matrices. Using this geometric perspective, it is easy to know that any unit vector in the plane spanned
by the two unit vectors represented by $\mathcal{C}_{1}(\bk_{\parallel})$ and $\mathcal{C}_{2}(\bk_{\parallel})$ satisfies
the anticommutation relation with the Hamiltonian. This fact suggests that the sub-chiral symmetry operator has
infinitely many choices. The existence of infinitely many choices just suggests that we have infinitely many ways
to understand the properties of the boundary states for such cases. This is quite different from
the conventional chiral symmetry whose operator has a definite form for a given Hamiltonian. This big
difference suggests that, compared to chiral-symmetric systems, sub-chiral symmetric
systems can have much richer topological properties. This big difference is worthy of in-depth studies,
and we leave them to future.

To conclude, from the detailed discussions above, it is clear that we can always choose a form for the sub-chiral symmetry operator
with only two $\gamma$ matrices. This is equivalent to the statement that we can always choose the sub-chiral symmetry operator
to have chiral symmetry.

\section{Quantum-material and metamaterial realization of sub-chiral symmetric systems\label{Appendixc}}

In the last section, we have elucidated that, if the ratio of two (or even more) terms of a Dirac Hamiltonian only
depends on partial components of the momentum, then the Hamiltonian has sub-chiral symmetry even
though the Hamiltonian does not have chiral symmetry. This is in fact a quite loose condition. As we will illustrate
below, there are many systems in which such a condition can be fulfilled.

In condensed matter systems, the chiral symmetry can be a sublattice symmetry or a relatively more abstract one
which corresponds to the product of time-reversal symmetry and particle-hole symmetry. Generally speaking,
the sublattice symmetry is an approximate symmetry in real quantum materials since the couplings within the same set of sublattices have no reason
to vanish exactly. In contrast, the abstract one can be an exact symmetry in the context of superconductors. This is because
a superconductor, as long as its description falls inside
the Bogoliubov-de Gennes framework, naturally has particle-hole symmetry. Then as long as the time-reversal symmetry is conserved,
the chiral symmetry is present.

Let us first consider the non-superconducting case. We are going to show that when the orbitals of the energy bands are appropriately chosen,
the aforementioned condition can  be fulfilled. There are many choices for the orbitals to satisfy our purpose. As a concrete example for illustration, let us assume that there are only two energy bands near the Fermi energy, with the orbital nature
of one band to be an $s$ orbital, and the orbital nature of the other band to be a $d_{xz}-id_{yz}$ orbital (the two orbitals play the role as a pseudospin). Let us further assume that a band inversion occurs between the two bands. According to the symmetry properties
of the orbitals, the effective tight-binding Hamiltonian needs to take the generic form
\begin{widetext}
\begin{eqnarray}
\mathcal{H}(\bk)=\left(
                   \begin{array}{cc}
                     m-t_{s}\cos k_x-t_{s}\cos k_{y}-t_{s}\cos k_{z} & \lambda\sin k_{z}(\sin k_{x}-i\sin k_{y}) \\
                     \lambda\sin k_{z}(\sin k_{x}+i\sin k_{y}) & -m+t_{d}\cos k_x+t_{d}\cos k_{y}+t_{dz}\cos k_{z}\\
                   \end{array}
                 \right)
\end{eqnarray}
\end{widetext}
Rewrite the Hamiltonian via the identity matrix and Pauli matrices,
\begin{eqnarray}
\mathcal{H}(\bk)&=&\left[\frac{\delta t}{2}(\cos k_{x}+\cos k_{y})+\frac{\delta t_{z}}{2}\cos k_{z}\right]\sigma_{0}\nonumber\\
&&+(m-\bar{t}\cos k_{x}-\bar{t}\cos k_{y}-\bar{t}_{z}\cos k_{z})\sigma_{z}\nonumber\\
&&+ \lambda\sin k_{z}\sin k_{x}\sigma_{x}+\lambda\sin k_{z}\sin k_{y}\sigma_{y},
\end{eqnarray}
where $\delta t=(t_{d}-t_{s})$, $\delta t_{z}=t_{dz}-t_{s}$, $\bar{t}=(t_{s}+t_{d})/2$ and $\bar{t}_{z}=(t_{s}+t_{dz})/2$.
The first term associated with the identity matrix makes the conduction and valence bands asymmetry, 
however, it does not affect the 
wave functions and hence is commonly neglected when one is only interested in the 
quantum geometry or topological properties of the wave functions. Without the first term, 
the Hamiltonian reduces as
\begin{eqnarray}
\mathcal{H}(\bk)&\simeq&[m-\bar{t}(\cos k_{x}+\cos k_{y})-\bar{t}_{z}\cos k_{z}]\sigma_{z}\nonumber\\
&&+\lambda\sin k_{z}\sin k_{x}\sigma_{x}+\lambda\sin k_{z}\sin k_{y}\sigma_{y}.
\end{eqnarray}
Apparently, this reduced Hamiltonian has sub-chiral symmetry, with the symmetry operator given by
\begin{eqnarray}
\mathcal{C}(k_{x},k_{y})=-\sin\theta(k_{x},k_{y}) \sigma_{y}+\cos\theta(k_{x},k_{y})\sigma_{x},
\end{eqnarray}
where $\theta(k_{x},k_{y})=\arg[\sin k_{x}+i\sin k_{y}]$. Similar to the three-dimensional Hamiltonian in the main text,
this Hamiltonian supports a mirror-symmetry-protected nodal ring in the bulk and a flat band on the top or bottom $z$-normal surface when the parameters are appropriately chosen, e.g., $2\bar{t}-\bar{t}_{z}<m<2\bar{t}+\bar{t}_{z}$. However, a big difference between them
is that, for here the three-dimensional Hamiltonian, the winding number characterizing the sub-chiral symmetry operator is equal to one.
Accordingly, if one considers a closed contour enclosing the surface time-reversal invariant momentum within the
projected region of the bulk nodal ring, then the pseudo-spin textures of the surface flat bands on this contour will display
a complete cycle of winding, and the Berry phase is $\pi$ (mod $2\pi$).

For the above example, the term associated with the identity matrix is in general finite in quantum material. Such a term
does not affect the existence and topological properties of the boundary flat bands, and its main effect 
to the boundary bands is to make their band width finite.
Nevertheless, if the band width of the surface bands is much smaller than other energy scales,
such as the band energy gaps, the surface bands remain an ideal platform
to investigate interaction/disorder driven phenomena.

Next let us discuss the situation in superconductors. When a superconductor hosts appropriate pairings, the superconductor
can also have sub-chiral symmetry. A standard classification of the pairings in a superconductor is in terms
of the irreducible representation of the symmetry group. Here for illustration we directly use the results presented in
the classic review paper~\cite{Sigrist1991}. Without loss of generality, let us focus on the symmetry group
$D_{4h}$ for illustration. For this symmetry group, the spin-singlet pairing corresponding to the irreducible representation $\Gamma_{5}^{+}$
takes the from $k_{z}(k_{x}+ik_{y})$ (see Table IV on page 9 and Table VI on page 11 in Ref.\cite{Sigrist1991}). If there is only one pair of spin-degenerate bands
at the Fermi energy, then the tight-binding Hamiltonian describing the superconductor with such a pairing, under the basis $\psi_{\bk}=(c_{\bk,\uparrow},c_{\bk,\downarrow},c_{-\bk,\uparrow}^{\dag},c_{-\bk,\downarrow}^{\dag})^{T}$,
is of the form
\begin{widetext}
\begin{eqnarray}
\mathcal{H}_{BdG}(\bk)=\left(
                         \begin{array}{cccc}
                           \epsilon(\bk)-\mu & 0 & 0 & \Delta(k_{x},k_{y}) \sin k_{z} \\
                           0 & \epsilon(\bk)-\mu & -\Delta(k_{x},k_{y}) \sin k_{z} & 0 \\
                           0 & -\Delta^{*}(k_{x},k_{y}) \sin k_{z} & -\epsilon(\bk)+\mu & 0 \\
                           \Delta^{*}(k_{x},k_{y}) \sin k_{z} & 0 & 0 & -\epsilon(\bk)+\mu \\
                         \end{array}
                       \right),
\end{eqnarray}
\end{widetext}
where $\epsilon(\bk)=-t(\cos k_{x}+\cos k_{y})-t_{z}\cos k_{z}$ describes the normal-state band structure,
$\mu$ the chemical potential, and $\Delta(k_{x},k_{y})=\Delta(\sin k_{x}+i\sin k_{y})$ the pairing gap function with $\Delta$ a real
constant. Because of the spin-rotation symmetry due to the absence of spin-orbit coupling, it is
easy to see that the Hamiltonian can be decoupled into two parts by a rearrangement of the basis. To be explicit,
by a change of the basis,
$(c_{\bk,\uparrow},c_{\bk,\downarrow},c_{-\bk,\uparrow}^{\dag},
c_{-\bk,\downarrow}^{\dag})^{T}\rightarrow(c_{\bk,\uparrow},c_{-\bk,\downarrow}^{\dag},c_{\bk,\downarrow},
c_{-\bk,\uparrow}^{\dag})^{T}$, the Hamiltonian becomes
\begin{widetext}
\begin{eqnarray}
\mathcal{H}_{BdG}(\bk)&=&\left(
                         \begin{array}{cccc}
                           \epsilon(\bk)-\mu & \Delta(k_{x},k_{y}) \sin k_{z} & 0 & 0 \\
                           \Delta^{*}(k_{x},k_{y}) \sin k_{z} & -\epsilon(\bk)+\mu & 0 & 0 \\
                           0 & 0 & \epsilon(\bk)-\mu & -\Delta(k_{x},k_{y}) \sin k_{z} \\
                           0 & 0 & -\Delta^{*}(k_{x},k_{y}) \sin k_{z} & -\epsilon(\bk)+\mu \\
                         \end{array}
                       \right)\nonumber\\
                       &=&\mathcal{H}_{u}(\bk)\oplus\mathcal{H}_{d}(\bk).
\end{eqnarray}
\end{widetext}
Let us focus on the upper two-by-two block, $\mathcal{H}_{u}(\bk)$. In terms of the Pauli matrices,
its explicit form is
\begin{eqnarray}
\mathcal{H}_{u}(\bk)&=&(\epsilon(\bk)-\mu)\sigma_{z}+\Delta\sin k_{z}\sin k_{x}\sigma_{x}\nonumber\\
&&-\Delta\sin k_{z}\sin k_{y}\sigma_{y}.
\end{eqnarray}
Apparently, the form of the sub-chiral symmetry operator is similar to the above case. However, here a fundamental difference
is that the absence of terms associated with the identity matrix does not require any approximation.
In other words, for here the superconducting case, the sub-chiral symmetry
is exact, and thereby the flatness of the boundary bands is also exact.

Last but not the least, let us give a discussion about the metamaterial realization of sub-chiral symmetric systems.
Compared to quantum materials, metamaterials are much easier to engineer the desired tight-binding Hamiltonians since the
hopping patterns can be designed as will. Here we take the circuit systems as an example for illustration. In
Ref.~\cite{Wu2022circuit}, Yu's group (one author of the current paper) have systematically designed
a series of  building blocks that can realize all kinds of two-by-two Pauli matrices.
By arranging these building blocks to form a periodic lattice and connecting them via conducting wires,
one can in principle realize Hamiltonians with very complicated matrix elements. For instance,
in a recent work, a three-dimensional two-band Hopf insulator, which involve much more complicated hopping patterns compared
to the two models concerned in the current paper, has been implemented by using these building blocks~\cite{Wang2023hopf}.
The two two-band models concerned in the current paper, and other two-band generalizations, can also
similarly be implemented via these building blocks.

\bibliography{dirac}

\end{document}